\documentstyle[12pt]{article}

\newcommand{\Aa}{A^{{\rm adv}}}
\newcommand{\Ao}{A^{{\rm out}}}
\newcommand{\be}{\begin{equation}}
\newcommand{\B}{B^0}
\newcommand{\Bc}{\bar{B}}
\newcommand{\con}{{\rm const.}}
\newcommand{\Ci}{{\cal C}^\infty}
\newcommand{\ee}{\end{equation}}
\newcommand{\I}{{\bf 1}}
\newcommand{\sn}{{\cal R}}
\newcommand{\sa}{{\cal B}}
\newcommand{\R}{{\rm {\bf R}}}
\newcommand{\Z}{{\rm {\bf Z}}}
\newcommand{\N}{{\rm {\bf N}}}
\newcommand{\C}{{\rm {\bf C}}}
\newcommand{\Se}{{\cal S}at_e}
\newcommand{\Sp}{{\cal S}at_{-e}}
\newcommand{\Sw}{{\cal R}ad}
\newcommand{\Sj}{{\cal S}(H, {\rm {\bf C}})}
\newcommand{\Sc}{{\cal S}(H, {\rm {\bf C}}{}^4)}
\newcommand{\St}{{\cal S}}
\newcommand{\bul}{\bullet}
\newcommand{\K}{{\cal K}}
\newcommand{\Ka}{{\cal K}\otimes_{{\rm alg}}{\cal K}}
\newcommand{\Fa}{{\cal F}}
\newcommand{\Op}{{\cal O}}
\newcommand{\Ob}{{\cal A}}
\newcommand{\Id}{{\cal I}}
\newcommand{\lb}{\left[}
\newcommand{\rb}{\right]}
\newcommand{\lp}{\left(}
\newcommand{\rp}{\right)}
\newcommand{\s}{\!\cdot\!}
\newcommand{\al}{\alpha}
\newcommand{\ac}{\overline{\alpha}}
\newcommand{\bt}{\beta}
\newcommand{\D}{\partial}
\newcommand{\dl}{d^2 l}
\newcommand{\df}{:=}
\newcommand{\dis}{\displaystyle}
\newcommand{\k}{\kappa}
\newcommand{\e}{\epsilon}
\newcommand{\ph}{\varphi}
\newcommand{\fc}{\overline{f}}

\newcommand{\F}{\Phi}
\newcommand{\g}{\gamma}
\newcommand{\G}{\Gamma}
\newcommand{\z}{\zeta}
\newcommand{\de}{\delta}
\newcommand{\m}{d\mu}
\newcommand{\n}{\nabla}
\newcommand{\oc}{\bar{o}}

\newcommand{\dV}{\dot{V}}
\newcommand{\Vo}{V^{{\rm out}}}
\newcommand{\dVo}{\dot{V}^{{\rm out}}}
\newcommand{\qV}{V^{{\rm op}}}
\newcommand{\p}{\psi}
\newcommand{\ch}{\chi}
\newcommand{\la}{\lambda}
\newcommand{\ov}{\overline}
\newcommand{\Hi}{{\cal H}}
\newcommand{\si}{\sigma}
\newcommand{\txt}{\textstyle}

\newtheorem{pr}{Proposition}
\newtheorem{lem}[pr]{Lemma}

\title{Asymptotic algebra for charged particles and radiation}
\author{${\rm A{\scriptstyle NDRZEJ}~H{\scriptstyle ERDEGEN}}$
\thanks{\mbox{Alexander von Humboldt Fellow; present address:} 
\mbox{Institute of Physics, Jagellonian University, Reymonta 4, 
30-059 Cracow, 
Poland}; \mbox{e-mail: herdegen@thrisc.if.uj.edu.pl}}\\ {\it II.
Institute of Theoretical Physics, Hamburg University,}\\ {\it Luruper
Chaussee 149, 22761 Hamburg, Germany}}
\date{}

\begin{document}
\maketitle
\vskip .3in
\begin{abstract}
\noindent

A $C^*$-algebra of asymptotic fields which properly describes the
infrared structure in quantum electrodynamics is proposed. The
algebra is generated by the null asymptotic of electromagnetic field
and the time asymptotic of charged matter fields which incorporate
the corresponding Coulomb fields. As a consequence Gauss' law is
satisfied in the algebraic setting.  Within this algebra the
observables can be identified by the principle of gauge invariance. A
class of representations of the asymptotic algebra is constructed
which resembles the Kulish-Faddeev treatment of electrically charged
asymptotic fields.
\hspace*{\fill}
\end{abstract}
\vfill
\eject

\begin{sloppypar}
\renewcommand{\thesection}{\arabic{section}}
\renewcommand{\theequation}{\arabic{section}.\arabic{equation}}
\renewcommand{\thepr}{\arabic{section}.\arabic{pr}}

\section{Introduction}
\label{int}

It is frequently stated that the excellent experimental confirmation
of quantum electrodynamics is not matched by sufficient understanding
of its theoretical foundations yet. Indeed, the need for better
understanding of the theory seems to be confirmed by steady
fundamental research. One could even point out that experimental
verification has to be considered as provisory, as long as the theory
has no completely firm status. This refers not only to the existing
experimental evidence.  New experimental arrangements may be needed
to test the results of further theoretical investigations. 

This may especially be the case in relation to the problems connected
with the long-range character of the electromagnetic interaction,
Gauss' law and the proper description of charged states.  These
problems manifest themselves in the infrared divergencies of
perturbational QED \cite{jau76}, the structure of uncountably many
superselection sectors \cite{zwa76,buch82}, the infraparticle problem
\cite{schr63,buch86} and the spontaneous breaking of the Lorentz group 
in charged representations of local observables \cite{froe79,buch86}.
All these questions have been investigated in various theoretical
set-ups, with varying emphasis on mathematical rigour on the one
hand, and concrete calculations on the other. An important step in
the developement of understanding of the long-range structure was the
realization, that it is the timelike resp. lightlike asymptotic
structure that is relevant here \cite{chu65,kib68,kul70,zwa76}. The work
of Kulish and Faddeev deserves special mentioning, as it generalized
Dollard's idea \cite{dol64} of asymptotic dynamics to the Gupta-Bleuler
formulation of quantum electrodynamics. A more careful analysis of
asymptotical charged states within this formalism was given by
Morchio and Strocchi \cite{mor83}.  In rigorous mathematical terms,
within the algebraic framework in quantum field theory, the
asymptotic electromagnetic field has been obtained as an LSZ-type
limit by Buchholz \cite{buch82}.  For reviews see the book by Jauch and
Rohrlich \cite{jau76}, the lecture notes by Morchio and Strocchi
\cite{mor85}, and the book by Haag \cite{haa92}.

In spite of the progress brought by these works, our understanding of
what an electron is, is still not very concrete. So far there do only
exist abstract, though rigorous, characterizations of electrically
charged particles \cite{fms79,bps91}. Even more importantly, the
concrete algebraic structure of the asymptotic quantum fields is
still unclear.  Thus investigations of properties such as the
spontaneous breaking of the Lorentz group in charged sectors or the
nonexistence of sharp masses (infraparticles) are frequently based on
ad hoc assumptions. Most general results are therefore of the "no-go"
type (see e.g. \cite{buch86}).

It is the aim of the present work to propose a concrete formulation
of the algebraic structure of asymptotic fields (and observables) of
electrons interacting with radiation.  This algebra resembles various
elements of standard knowledge on the infrared problem. However, we
would like to point out three novel aspects of our formulation. \\
(i) A clear and consistent algebraic framework is obtained.  The
asymptotic algebra is a $C^*$-algebra, thereby placing the problem on
firm mathematical grounds.\\ (ii) We do not have to consider the
fully interacting quantum field theory.  In particular, the canonical
quantization (which is taken for granted in the treatment of Kulish
and Faddeev) is avoided. In fact, this heuristic procedure for
obtaining the algebraic structure may fail in the field of long-range
problems, as we shall discuss. Instead we base our construction on
the asymptotic structure of classical electrodynamics (Maxwell and
Dirac) which has been established in \cite{her95}.  These results lead
naturally via the correspondence principle to our quantum algebra.\\
(iii) The algebra incorporates the Coulomb fields of the asymptotic
(outgoing) particles. In this respect it resembles the theory of the
quantized Coulomb field of Staruszkiewicz \cite{sta}. However, the
latter is an idealized theory of certain isolated degrees of freedom,
and it seems to have no natural embedding into a larger scheme. 

The plan of the paper is as follows. In Sec.\ref{clas} we describe
briefly the asymptotic structure of the classical theory. Part of the
material is shifted to the Appendix. In Sec.\ref{quant} this
structure is then quantized according to the correspondence
principle. Heuristical physical considerations are presented which
lead to formal algebraic relations for the fields. These relations
are made mathematically precise in Sec.\ref{alg}, and lead to our
asymptotic $C^*$-algebra.  A class of representations of the algebra
is constructed in Sec.\ref{rep}. The question of physical relevance
of these representations is left for future work. Sec.\ref{dis}
brings some final remarks, and comments on future perspectives of
this work.

\section{Classical asymptotic structure}
\label{clas}

In this section we give a short account of the asymptotic structure
of a classical field theory with electromagnetic interaction,
discussed at length in \cite{her95}. For the definiteness we considered
the Dirac field interacting with radiation. Complete rigorous results
were obtained along the lines summarized here for the both external
field problems, but the extension to the full theory is possible, as
argued in \cite{her95}, under plausible conjectures. (The relation of
our asymptotic variables to those used by Flato et al. \cite{fla} 
in their recent solution of the Cauchy problem and proof of 
asymptotic completeness of the Maxwell-Dirac system is an open 
problem.) The aim of this
summary is also, for the convenience of the reader, to rewrite in the
tensor form the properties and formulas for the asymptotics of the
electromagnetic field which were discussed in the two-component
spinor language in \cite{her95}. Equivalence to the original formulation
is proved in the Appendix.

The idea of the approach deviates from the standard formulation of
the scattering problem as the limit of constant time configurations
with time tending to infinity. Rather, the advantage is taken of the
different propagation velocities of matter and radiation, to consider
their asymptotics in different spacetime regions. For the 
electromagnetic field the known methods of the null infinity
asymptotics \cite{bra77,ash86} are applied, formulated in terms of
homogeneous functions (without the Penrose's spacetime
compactification) and further developed in some specific aspects. For
the matter field a method is developed which leads to the
determination of an asymptotic field inside the forward lightcone --
this is sufficient, as eventually every massive particle enters the
cone.  Plausible arguments then indicate that asymptotics thus
defined contain the full information on the system and the total
Poincar\'e quantities (energy-momentum and four-dimensional angular
momentum) of the theory may be expressed in terms of them.

Electromagnetic field of the system admits a class of gauges of the
potential with the null asymptotic of the form (Eq.(2.45) in
\cite{her95})
\be
\lim_{R\to\infty}R\, A_a(x + Rl) = V_a(x\s l, l)\, .
\label{null}
\ee
Here $x$ is any spacetime point in Minkowski space $M$ and $l$ is a
future-pointing null vector. The function $V_a(s, l)$ ($s$ is a real
variable) is homogeneous of degree $-1$
\be
V_a(\mu s,\mu l) = \mu^{-1} V_a(s, l)\, ,
\label{hom}
\ee
($\mu>0$), and satisfies
\be
l\s V(s, l) = Q\, ,
\label{Q}
\ee
where $Q$ is the charge of the field. Its $s$-derivative
$\dis\dV_a(s, l)\df \frac{\D}{\D s} V_a(s, l)$ falls off according to
\be
|\dV_a(s, l)|<\frac{\con}{(1+|s|)^{1+\e}}
\label{dec}
\ee
for some $\e>0$, so $V_a(s, l)$ has limits for $s\to\pm\infty$, which
we denote $V_a(\pm\infty, l)$. (Null vectors $l$ are scaled in
(\ref{dec}) to $l^0=1$ in arbitrarily chosen, fixed Minkowski frame;
change of frame results only in change of bounding constant.) Gauge
freedom consists of the transformation $V_a(s, l)\to V_a(s, l) +
\al(s, l) l_a$. 

Further properties of the limit values $V_a(\pm\infty, l)$ involve
differentiation on cone variables. A simple and explicitely
Lorentz-covariant way to express the differentiations in directions
tangent to the cone is to use the operator $l_a\D_b - l_b\D_a$
($\D_a\df\D/\D l^a$). One applies the operator to any differentiable
extension to some neighbourhood of the cone of a function defined on
the cone itself, and restricts the result again to the cone. The
result is independent of the extension used. 

The limit values $V_a(\pm\infty, l)$ are constrained by the following
gauge-invariant condition
\be
l_{[a}\D_b V_{c]}(\pm\infty, l) = 0
\label{cond}
\ee
(this is the tensor form of the conditions (3.32) with (2.54) --
(2.57) in \cite{her95}; see also the Appendix). The physical content of the
condition, which is satisfied in standard scattering situations, is,
that it allows the identification of the total angular momentum of
the system, as discussed in \cite{her95}. The properties (\ref{hom}),
(\ref{Q}) and (\ref{cond}) allow another simple representation of
$V_a(\pm\infty, l)$. It follows from (\ref{hom}) and (\ref{Q}) that
$(l_b\D_a - l_a\D_b)V^b(\pm\infty, l) + V_a(\pm\infty, l) \propto
l_a$.  We denote 
\be
(l_b\D_a - l_a\D_b)V^b(+\infty, l) + V_a(+\infty, l) = 2 q(l) l_a\, ,
\label{q}
\ee
\be
(l_b\D_a - l_a\D_b)V^b(-\infty, l) + V_a(-\infty, l) = 2 \k(l) l_a\,
. 
\label{k}
\ee
$q(l)$ and $\k(l)$ are homogeneous functions of degree $-2$. If, for
the sake of differentiation, the extensions of $V_a(\pm\infty, l)$
are chosen so as to satisfy (\ref{Q}) also in a neighbourhood of the
cone, then
\be
q(l) = - \frac{1}{2} \D_b V^b(+\infty, l)\, ~~~
\k(l) = = - \frac{1}{2} \D_b V^b(-\infty, l)\, .
\label{kuk}
\ee
The charge of the field is recovered from the functions $q(l)$ and
$\k(l)$ by the formulas
\be
Q= \frac{1}{2\pi}\int q(l)\dl = \frac{1}{2\pi}\int \k(l)\dl\, .
\label{Q2}
\ee
By $\dl$ we denote that measure on the set of null directions which
gives a Lorentz invariant result when applied to a homogeneous
function of $l$ of degree $-2$ (see \cite{her95} and references given
there).  The measure itself is homogeneous of degree $2$, and for
$l$'s scaled to $l^0 = 1$ in any fixed Minkowski frame it is the
rotationally invariant measure on the unit sphere of vectors
$\vec{l}$.  For future reference we note that if $\al(l)$ is
differentiable and homogeneous of degree $-2$, then (see the Appendix) 
\be
\int (l_a\D_b - l_b\D_a) \al(l) \dl = 0\, .
\label{ip}
\ee

Now conversely, one can show that if $q(l)$ and $\k(l)$ are
homogeneous functions of degree $-2$ satisfying (\ref{Q2}), then the
vector functions $V_a(\pm\infty, l)$ constrained by (\ref{hom}),
(\ref{Q}) and (\ref{cond}) are determined by (\ref{q}) and (\ref{k})
uniquely up to a gauge (\cite{her95}, after Eq. (2.61)). 

Physical interpretation of $\k(l)$ is of importance: this function
determines the flux of the electromagnetic field in spacelike
infinity. Explicitely, if $x$ is any point and $y$ a spacelike
vector, then
\be
\lim_{R\to\infty} R^2 F_{ab}(x+Ry) = K_a(y) y_b - K_b(y) y_a\, ,
\label{fl1}
\ee
where
\be
K_a(y) = \frac{1}{2\pi y^2} \n_a\int \k(l)\, {\rm sgn}\, y\s l\, \dl\, .
\label{fl2}
\ee

If the potential is decomposed in standard way into the advanced and
the free outgoing part $A_a = \Aa{}_a + \Ao{}_a$, then $\Aa{}_a$ and
$\Ao{}_a$ have again asymptotics of the type (\ref{null}) with
$V_a(+\infty, l)$ and $\Vo{}_a(s, l) \df V_a(s, l) - V_a(+\infty, l)$
replacing $V_a(s, l)$ on the right hand side (rhs) of (\ref{null})
respectively. This brings the physical interpretation of $q(l)$: this
function determines the asymptotic Coulomb field of the outgoing
currents.

The free outgoing field is completely determined by its asymptotic
according to the formula
\be
\Ao{}_a(x)=-\frac{1}{2\pi}\int \dot{V}^{{\rm out}}{}_a(x\s l, l)\, \dl\, .
\label{kir}
\ee
The flux of the field $F^{{\rm out}}{}_{ab}(x)$ at the spacelike
infinity is again given by the formulas (\ref{fl1}) and (\ref{fl2})
in which $\k(l)$ is replaced by $\si(l)\df \k(l) - q(l) =
-\frac{1}{2}
\D^b\Vo{}_b(-\infty, l)$, which is therefore interpreted as the infrared 
characteristic of the free field. The vector function
$\Vo{}_a(-\infty, l)$ is again determined up to a gauge by $\si(l)$,
but in this case a more explicit representation is possible. One
shows that the equation (\ref{cond}) (satisfied by $\Vo{}_a(-\infty,
l)$) and the condition $l\s \Vo(-\infty, l)= 0$ together imply the
existence of a real function $\F(l)$ homogeneous of degree $0$ such
that (see the Appendix)
\be
l_{[a}\Vo{}_{b]}(-\infty, l) = l_{[a}\D_{b]}\F(l)\, .
\label{gra}
\ee
$\F(l)$ is determined by this condition up to an additive constant.
We make the choice of this constant characterize classes of gauges of
$\Vo{}_a(-\infty, l)$ as follows. First, fix $\F(l)$ with some choice
of the constant. Then, choose an antisymmetric real tensor function
$G_{ab}(l) = -G_{ba}(l)$, homogeneous of degree~$0$, such that
\be
l_{[a}G_{bc]}(l) = 0\, ,~~~G_{ab}(l)\, l^b = \F(l)\, l_a\, .
\label{G}
\ee
This tensor is then of the form $G_{ab}(l) = l_a g_b(l) - l_b
g_a(l)$, where $g_a(l)$ is homogeneous of degree $-1$, satisfies 
\be
l\s g(l) = \F(l)\, ,
\label{lg}
\ee
and is determined by $G_{ab}(l)$ up to $g_a(l)\to g_a(l) + \al(l)
l_a$.  Finally, put 
\be
\Vo{}_a(-\infty, l) = (l_b\D_a - l_a\D_b) g^b(l) + g_a(l)\, .
\label{g}
\ee
One shows that\\ (i) (\ref{g}) satisfies (\ref{gra});\\ (ii) every
gauge of $\Vo{}_a(-\infty, l)$ may be represented in this way and the
correspondence with the tensors $G_{ab}(l)$ is $1:1$;\\ (iii) gauges
corresponding to the choices of $G_{ab}$'s with a fixed constant in
$\F(l)$ form equivalence classes with respect to the following
equivalence relation: $\dis\Vo_1{}_a(-\infty, l)\sim
\Vo_2{}_a(-\infty, l)$
iff $\dis\int \al(l)\,\dl = 0$, where $\dis\Vo_1{}_a(-\infty, l)-
\Vo_2{}_a(-\infty, l) = \al(l) l_a$. We shall see in 
the next section that the interpretation imposed by the above
procedure on the additive constant in $\F(l)$ is a very natural one
from the point of view of the symplectic form for electromagnetic
field.

The asymptotic of the Dirac field $\psi(x)$ inside the forward
lightcone is determined by considering the behaviour of $\psi(\la v)$
for large $\la$, where the four-velocity $v$ lies on the hyperboloid
$H\df \{ v\in M|\, v^2=1, v^0>0\}$.  Physically, the most important
condition for the validity of our discussion is such a choice of
gauge of the electromagnetic potential, that $|v\s A(\la v)|<
\con\la^{-1-\e}$, ($\e>0$). That this is possible in our 
context is shown in \cite{her95}, where further details are also given.
One shows then, that if we put $$
\p(\la v) = -i\, \la^{-3/2}\, e^{\dis -i(m\la + \pi /4)\g\s v}
f_\la (v) $$ ($\g^a$ are the Dirac matrices), then
$\lim_{\la\to\infty} f_\la(v) = f(v)$. For the external field problem
it is shown that this limit is reached strongly in the Hilbert space
$\K$ of four-component functions on the hyperboloid $H$ with the
scalar product
\be
(f, g) = \int \ov{f(v)}\, \g\s v\, g(v)\, \m(v)\, ,
\label{sp}
\ee
where bar denotes the usual Dirac conjugation and $\m(v)=d^3 v/v^0$
is the invariant measure on the hyperboloid. Moreover, free outgoing
field may be determined by 
\be
\psi^{{\rm out}}(x) = \lp\frac{m}{2\pi}\rp^{3/2} 
\int e^{{\textstyle-im\,x \s v\,\g\s v}} \g\s v\, f(v)\,\m(v)\,  .
\label{dfr}
\ee
$\psi^{{\rm out}}(x)$ has the same asymptotic as $\psi(x)$. It is
argued that the structure is essentially the same in the full theory.
In that case the characteristic $q(l)$ of the Coulomb field of the
outgoing currents is expressed in terms of their asymptotic
\be
q(l) = e \int \ov{f(v)}\, \g\s v\, f(v)\, \frac{\m(v)}{2(v\s l)^2}\,
,
\label{mc}
\ee
($e$ is the charge of the electron).

We note for future use that explicitely Lorentz-covariant
differentiation on $H$ may be discussed with the use of operators
defined in exactly the same way as it has been done in the case of
the lightcone in the paragraph preceding (\ref{cond}). In the case of
hyperboloid the operator contracted with $v^b$ contains the whole
information
\be
\de_af(v)\df (\D_a - v_a v\s \D) f(v)\, .
\label{dif}
\ee

Finally, total Poincar\'e quantities are expressed in terms of the
outgoing fields. One finds that these quantities are the sums of the
respective quantities for the free fields $F^{{\rm out}}{}_{ab}(x)$
and $\psi^{{\rm out}}(x)$ (deteremined by (\ref{kir}) and (\ref{dfr})
respectively), with, however, one additional term in the case of
angular momentum
\be
\Delta M_{ab} = - \frac{1}{2\pi} \int q(l) (l_a\D_b - l_b\D_a)\F(l)\, \dl\, .
\label{am}
\ee
This term is seen to mix the outgoing matter characteristic $q(l)$
with the infrared characteristic $\F(l)$. (It originates from the
mixed adv-out terms of the asymptotics in null directions of the
electromagnetic energy-momentum tensor.) This mixing of the
long-range degrees of freedom corresponds to this remnant of
interaction which is responsible for the validity of the Gauss' law.
Its appearence shows, that a Poisson bracket structure separating the
two fields asymptotically remains in contradiction not only with the
Gauss' law, but also with the Poincar\'e structure of the theory.

\setcounter{equation}{0}
\setcounter{pr}{0}

\section{Quantization}
\label{quant}

We assume now that the asymptotic structure of the quantum theory may
be described by quantum variables analogous to $V_a(s, l)$ and
$f(v)$.  By that we mean, that these analogs generate an algebra, the
states of which may be interpreted as scattering states in quantum
electrodynamics.  In the present section we give heuristic arguments
which lead us to the formulation of quantization conditions for these
variables. In the next section then the appropriate algebra is
constructed.

The usual quantization of the free electromagnetic field is achieved
by the use of the symplectic form
\be
\{ F_1, F_2\} = \frac{1}{4\pi} \int_{\Sigma} (F_1{}^{ab}A_2{}_b - 
F_2{}^{ab}A_1{}_b)\, d\si_a\, ,
\label{ss}
\ee
the integration extending over a Cauchy surface $\Sigma$.  It has
been observed by other authors before \cite{bra77,ash86} that the
integration surface may be shifted so as to become the future null
infinity hypersurface (in the language of the compactified Minkowski
space), as the fields are determined by their data on this surface.
This corresponds in our language (no compactification) to the
integration of "radiated" symplectic form.  This is calculated in
exactly the same way in which the radiated energy-momentum and
angular momentum were determined in \cite{her95}, Eq.(3.6--3.14).  The
explicitely Lorentz-invariant result, denoted by $\{ V_1, V_2\}$, is
\be
\{V_1, V_2\} = \frac{1}{4\pi}\int (\dV_1\s V_2 - \dV_2\s V_1)(s, l)\, 
ds\, \dl\, ,
\label{sn}
\ee
where $V_i$ are asymptotics (\ref{null}).  However, we observe that
also in the presence of sources the electromagnetic field is locally
free in the null asymptotic region, so one can try to use the same
symplectic form for the asymptotics of the interacting theory.  The
form (\ref{sn}) is now extended without formal change to asymptotics
of all fields admitted by the framework of Sec.\ref{clas}.  (The
reason for taking (\ref{sn}) rather then directly the "radiated"
analog of (\ref{ss}) for general fields as a basis for generalization
is that for charged fields with nonvanishing infrared part the latter
form yields no Lorentz invariant result. If the calculation is
performed in a frame with the time-axis along the positive unit
timelike vector $t$ then the result differs from (\ref{sn}) by $$
\frac{Q_2}{4\pi}\int t^a\Vo_1{}_a(-\infty, l)\, \frac{\dl}{t\s l} - 
\frac{Q_1}{4\pi}\int t^a\Vo_2{}_a(-\infty, l)\, \frac{\dl}{t\s l} \, ,
$$ where $Q_i$ are the charges (\ref{Q}).) 

If $V$'s are split into the free and the Coulomb part $V_a(s, l) =
\Vo{}_a (s, l) + V_a(+\infty, l)$ then 
$$
\begin{array}{@{}l}
\dis\{V_1, V_2\} = \{\Vo_1, \Vo_2\}\\[2mm] 
\dis+ \frac{1}{4\pi}\int 
(V_1(+\infty, l)\s \Vo_2(-\infty, l) - V_2(+\infty, l)\s
\Vo_1(-\infty, l) )\, \dl\, .
\end{array}
$$ Substituting in the second term (\ref{g}), integrating by parts
with the use of (\ref{ip}), and finally using (\ref{q}) and
(\ref{lg}) one obtains
\be
\begin{array}{@{}c}
\dis\{V_1, V_2\} = \frac{1}{4\pi}\int (\dVo_1\s \Vo_2 - \dVo_2\s \Vo_1)(s, l)\, 
ds\, \dl\\[2mm]
\dis+ \frac{1}{2\pi}\int (q_1\F_2 - q_2\F_1)(l)\, \dl\, .
\end{array}
\label{split}
\ee
The first term on the rhs is gauge invariant, while the second
depends on gauge only through the choice of the additive constant in
$\F(l)$, that is on the choice of one of the equivalence classes of
$\Vo{}_a(-\infty, l)$ discussed after Eq. (\ref{g}). The above
compact form of the second term supplies justification for our
interpretation of the constant in $\F(l)$.

Let now $\qV{}_a(s, l)$ be a quantum field and $V_a(s, l)$ a
classical test field. The heuristic quantization rule is 
$$
\left[ \{V_1, \qV\}, \{V_2, \qV\} \right] = i \{V_1, V_2\}\, ,
$$ 
where the real multiplicative constant on the rhs is fixed by the
condition, that the quantization reduces to the standard one for free
infrared-regular test fields. In the Weyl exponentiated form this
becomes $\dis W(V_1)W(V_2) = e^{-(i\bt^2/2)\{V_1, V_2\}} W(V_1+V_2)$,
where we put $\dis W(V)=e^{-i\bt\{V, \qV\}}$, with $\bt$ a real
constant to be determined shortly. The Weyl operators are assumed to
depend only on those variables, which enter nontrivially into the
symplectic form (\ref{split}), that is they are insensitive to the
gauge of $\Vo{}_a(s, l)$ for finite $s$ and to the gauge of
$V_a(+\infty, l)$, and they depend on the gauge of $\Vo{}_a(-\infty,
l)$ only through the choice of constant in $\F(l)$.  Therefore we
shall write sometimes $W(V) = W(\xi, \F, q)$, where $\xi_{ab}(s,
l)\df l_a\Vo{}_b(s, l) - l_b\Vo{}_a(s, l)$. (Remember that $\F(l)$ is
determined by $\xi_{ab}(-\infty, l)$ up to an additive constant.) The
form (\ref{split}) determines the physical interpretation of the Weyl
operator for $q=0$, $\xi=0$, $\F=c=\con$: $W(0, c, 0) = e^{\txt i\bt
cQ^{{\rm op}}}$, where $Q^{{\rm op}}$ is the operator of the charge of
the field. From Weyl relations we have then 
$$ 
e^{i\bt cQ^{{\rm op}}}
W(V) = W(V) e^{i\bt c(Q^{{\rm op}} + \bt Q)}\, , 
$$ 
where $Q$ is the
charge of the electromagnetic test field $V$. For the interpretation
of the classical and the quantum charge to agree we set $\bt=1$. Then
the Weyl operator $W(V)$ carries a quantum charge equal to the
classical charge of the test field $V$. More generally, it may be
interpreted to carry the asymptotic field characterized by $V_a(s,
l)$.  The Weyl algebra
\be
W(V_1)W(V_2) = e^{-(i/2)\{V_1, V_2\}} W(V_1+V_2)
\label{w}
\ee
may be considered as a theory of the asymptotic electromagnetic
field.  The quantization of charge demands that the space of test
fields be restricted to the abelian additive group of those $V$'s
which carry the multiple of the elementary charge $\dis
Q=(1/2\pi)\int q(l)\dl = ne$.  The subgroup of zero-charge test
fields forms a vector space.  (For the discussion of an "adiabatic
limit" of such a theory, in which only the long-range characteristics
of the field survive, we refer the reader to
\cite{her94}.) However, the theory thus formulated is physically incomplete -- 
it admits Coulomb fields, but there are no particles present to carry
these fields. We turn now to the description of these particles. 

Let us forget for the moment that there is some "Gauss coupling"
between the asymptotic electromagnetic and Dirac fields which has to
modify the Poisson bracket structure (as compared with the structure
of two independent fields). Then the quantum field $f^{{\rm op}}(v)$
which is to correspond to the classical $f(v)$ is quantized in the
standard Dirac way. We denote the quantum field smeared with the test
field $f(v)$ by $B(f)$ and replace $f^{{\rm op}}(v)$ by $B(v)$, so
that symbolically $\dis B(f) = \int \ov{f(v)}\, \g\s v\, B(v)\,
\m(v)$, where $f$ is in $\K$, 
the Hilbert space introduced before (\ref{sp}).  The standard
quantization law in our notation reads $$ [B(f), B(g)]_+ = 0\, ,~~~
[B(f), B(g)^*]_+ = (f, g)\, .  $$ It will be convenient for our
purposes to have fermionic operators depending on $f$'s linearly.  We
introduce notation
\be
\B(f) = B(f^c) = \int \B(v)\, \g\s v\, f(v)\, \m(v)\, ,
\label{B}
\ee
\be
\Bc(f) = B(f)^* = \int \Bc(v)\, \g\s v\, f(v)\, \m(v)\, ,
\label{Bc}
\ee
where $f^c$ is the charge conjugation of $f$ defined by
\be
f^c= C\fc^{{\rm T}}\label{cg}\, ,
\ee
with $C$ a unitary, antisymmetric matrix inducing the transformation
\\
$C^{-1}\g^aC=-\g^a{}^{{\rm T}}$. The involution law is then
\be
\B(f)^* =\Bc(f^c)\, ,
\label{conj}
\ee
and anticommutation relations
\be
[\B(f), \B(g)]_+ = 0\, ,~~~ [\B(f), \Bc(g)]_+ = (f^c, g)\, ,
\label{com1}
\ee
or, symbolically,
\be
[\B(v)_\al, \B(v)_\bt]_+ = 0\, ,~~~ [\B(v)_\al, \Bc(u)_\bt]_+ = \de(v, u)
(C^{-1}\g\s v)_{\al\bt}\, ,
\label{com2}
\ee
where $\de(v, u)$ is the Dirac "$\de$-function" in the two velocities
with respect to the measure $\m(v)$. 

Physical interpretation of $B(f)$ and $\B(f)$ justified by the Fock
representation is, that these operators annihilate an electron and/or
create a positron. This means that they locally create charge $-e$
(if $e$ is the charge of electron). More specifically, the operators
$B(v)$ and $\B(v)$ (resp. $\Bc(v)$) (forget for the moment
mathematical subtleties) create the charge $-e$ (resp. $e$) moving
with a constant four-velocity $v$.  However, if the Gauss' law is
again brought into play, creation of a charged particle must have
electromagnetic consequences. Therefore, we want to extend the effect
of $B$'s in such a way, that they create (or annihilate) also the
Coulomb field accompanying the charge.  Basing the intuitions on
pictures from perturbation calculations and on the algebraic
discussion of superselection sectors of local observables, we want to
admit the possibility, that charged particles are in addition
accompanied by "clouds" of radiation. Let the potential of the total
field (Coulomb + radiation) accompaning charge $e$ moving with
velocity $v$ be characterized by the asymptotic $V_a(v)=V_a(v; s,
l)$. Then for each $v$ this asymptotic is in the class discussed in
Sec.\ref{clas}, and, moreover, its Coulomb part is, up to a gauge,
the asymptotic of the potential $\dis A_a(x)=ev_a/v\s x$, that is
$\dis V_a(v; +\infty, l) = ev_a/v\s l + {{\rm gauge}}$. Now we seek
operators $\dis\B_{-V_1}(v)$ and $\dis\Bc_{V_2}(v)$ analogous to
$\B(v)$ and $\Bc(v)$ respectively, which, however, beside creating or
annihilating material particles should also carry accompaning
electromagnetic fields with asymptotics $-V_1(v)$ and $V_2(v)$
respectively, where $V_1$ and $V_2$ are in the class introduced
above.  Thus formulated, the problem almost uniquely determines its
solution -- the objects which do the electromagnetic part of the task
are already there, the Weyl operators $W(-V_1(v))$ and $W(V_2(v))$
carry exactly those charged fields. For the purpose of obtaining
commutation relations we imagine the operators $\B_{-V_1}(v)$ (resp.
$\Bc_{V_2}(v)$) to be formed as products of $\B(v)$ (resp. $\Bc(v)$)
and $W(-V_1(v))$ (resp. $W(V_2(v))$) (mutually commuting). On the
other hand, having attached charged fields to the matter particles in
this way, we do not need any longer, as independent objects, the Weyl
operators $W(V)$ for test fields with nonvanishing Coulomb part. From
now on the asymptotic $V$ in $W(V)$ is always a free field asymptotic
($V_a(+\infty, l)=0$).

The (naive) commutation relations resulting from the above discussion
are
\begin{eqnarray*}
&&\B_{-V_1}(v) W(V) = e^{(i/2)\{ V_1(v), V\}} \B_{-V_1 + V}(v)\,
,\\[3mm] &&\Bc_{V_2}(v) W(V) = e^{-(i/2)\{ V_2(v), V\}} \Bc_{V_2 +
V}(v)\, ,\\[3mm] &&e^{(i/2)\{ V_1(v), V_2(u)\}}\B_{-V_1}(v)_\al
\B_{-V_2}(u)_\bt 
+e^{(i/2)\{ V_2(u), V_1(v)\}}\B_{-V_2}(u)_\bt \B_{-V_1}(v)_\al = 0\,
,\\[3mm] &&e^{-(i/2)\{ V_1(v), V_2(u)\}}\B_{-V_1}(v)_\al
\Bc_{V_2}(u)_\bt 
+e^{-(i/2)\{ V_2(u), V_1(v)\}}\Bc_{V_2}(u)_\bt \B_{-V_1}(v)_\al \\[3mm]
&&\hspace*{2cm}= \de(v, u)(C^{-1}\g\s v)_{\al\bt}W(V_2(v) - V_1(v))\, ,
\end{eqnarray*}
supplemented with the Weyl relations. Note that the asymptotic in the
Weyl operator in the last line is a free field asymptotic, as $V_2(v;
+\infty, l) - V_1(v; +\infty, l) = 0$.

For a precise formulation of the above quantization conditions it
will not suffice, in contrast to the case of the Dirac field algebra, to
have objects smeared with one-particle test functions as generating
elements. Smeared products of $\dis\B_{-V_1}(v)$, $\dis\Bc_{V_2}(v)$
and $W(V_2(v) - V_1(v))$ have to be defined, and the phase factors
appearing in the above relations must become multipliers in the space
of test fields. This construction is given in the next section. 

Before we go over to this task we want to draw attention to a known
fact concerning physical interpretation of Weyl operators, which,
however, in our context is of decisive importance and must not be
overlooked. We illustrate our point first on the simplest possible
example, the Weyl algebra of a single pair of canonical variables in
quantum mechanics, $[x, p]=i\I$.  The Weyl formulation reads in that
case $$ W(x_1, p_1) W(x_2, p_2) = e^{-(i/2)\{ x_1, p_1; x_2, p_2\}}
W(x_1+x_2, p_1+p_2)\, , $$ where $x_i$ and $p_i$ are classical
"test"- position and momentum variables, the symplectic form is $\{
x_1, p_1; x_2, p_2\}= x_1p_2 - x_2p_1$, and the Weyl operator is
interpreted as $\dis W(x_1, p_1) = e^{-i\{ x_1, p_1; x, p\}}$. We
want to point out a certain duality in the interpretation of $W(x_1,
p_1)$. Let us set for simplicity $p_1=0$. Then, on the one hand, if
treated as a function of observables, the Weyl operator $\dis W(x_1,
0)=e^{-ix_1 p}$ "measures" the probability distribution of a state
with respect to the momentum. On the other hand we have $\dis W(x_1,
0)^* x\, W(x_1, 0) = x+x_1$. Hence, if treated as a unitary
transformation operator, $W(x_1, 0)$ "carries" the translation $x_1$,
that is when applied to a vector in Hilbert space it increases its
position characteristic by the "test"- position $x_1$. The
characteristic which is "carried" by a Weyl operator in this sense is
thus the one given by the test-quantity, while the "measured" one is
dual to it, in the sense of the symplectic form. Going back to our
objects we see that $\dis\B_{-V_1}(v)$, $\dis\Bc_{V_2}(v)$ and $W(V)$
carry the respective fields in the above sense, in the case of $W(V)$
a free field. However, from the symplectic form (\ref{split}) one
reads off, that the asymptotic of a free field potential has as its
dual the asymptotic of the total field, which is therefore what is
"measured" by $W(V)$.

\setcounter{equation}{0}
\setcounter{pr}{0}

\section{The algebra}
\label{alg}

Consider the set of $\Ci$ functions $V_a(s, l)$ (differentiations
with respect to $l$ in the sense discussed before Eq.(\ref{cond}),
outside some neighbourhood of the vertex of the cone) satisfying
conditions (\ref{hom} --
\ref{cond}). In this set introduce the following equivalence relation: 
$V_2(s, l)\sim V_1(s, l)$ iff $V_2(s, l)= V_1(s, l) + \al(s, l)\,l$
and $\dis\int\lp\al(-\infty, l) - \al(+\infty, l)\rp\, \dl=0$ (this is
the equivalence relation for the infrared characteristics of the free
field component of $V_a$, introduced in (iii) after (\ref{g})). The
set of equivalence classes with respect to this relation will be
denoted by $L_Q$.  Another way of characterizing elements of $L_Q$ is
by the triples $(\xi, \F, q)$ introduced in the paragraph preceding
Eq.(\ref{w}). In order not to burden the notation, the elements of
$L_Q$ will be denoted by $V_a(s, l)$, but always the equivalence
classes are understood.  The set $\dis\hat{L}\df\bigcup_{n\in\Z}
L_{ne}$ has in natural way the structure of an abelian additive
group. With the map $\{ .,.\}: \hat{L}\times\hat{L}\to\R$ defined by
(\ref{sn}) it becomes a symplectic group, on which $\{ .,.\}$ is
nondegenerate. The subgroup $L_0\subset\hat{L}$ has the structure of
a vector space.  Its subspace consisting of elements satisfying in
addition $V_a(+\infty, l)\propto l_a$ (no Coulomb field) will be
denoted by $L$. Without
loss of generality it may be assumed that for all $V$'s in $L$ there
is $V_a(+\infty, l)=0$. Elements of $L$ will be the test functions of
Weyl operators.

Consider, next, the class of all functions $V_a(v)$ on the
hyperboloid $H$ with values in $L_e$, such that \\[1mm] 
(i) $\dis
V^a(v; +\infty, l)= V_e{}^a(v, l) +{{\rm gauge}}$, where $\dis
V_e{}^a(v, l)\df ev^a/v\s l$, or, equivalently, $\dis q(v, l)= q_e(v,
l)\df (e/2)(v\s l)^{-2}$;\\[1mm] 
(ii) for each $V(v)$ there is a
function $F_V(v)$ such that
\be
\{ V(v), V(u)\} = F_V(v) - F_V(u)
\label{sep}
\ee
for every $v$ and $u$;\\[1mm] 
(iii) representants $V_a(v; s, l)$ may
be chosen in $\Ci$ in $v$ (differentiations in the sense of
(\ref{dif})) and for each $k=0,1,\ldots$ there are constants
$C_k\in\R$ and $m_k\in\N\cup\{ 0\}$ such that $$
\left|\de_{a_1}\ldots\de_{a_k} \dV_b(v; s, l)\right|< 
C_k(v^0)^{m_k}(|s|+1)^{-1-\e} $$ (in a fixed Minkowski frame, with
scaling of $l$'s fixed by $l^0=1$; the change of frame induces only a
change of $C_k$).\\[1mm] 
Let $\Se$ be a subclass of this family of
functions, such that \\[1mm] 
(iv) if $V_1(v)\in\Se$ and $V_0\in L$
then $V_2(v)=V_1(v) + V_0\in\Se$; this condition is fulfilled if
$\Se$ consists of all functions satisfying (i)--(iii). \\[1mm]
Denote, moreover, $\Sp\df-\Se$ and $\Sw\df\Se +\Sp$. The elements of
$\Se$ and $\Sp$ will be the fields accompanying particles. Free
fields from the class $\Sw$ will serve to define smeared Weyl
operators.

Physical meaning of the first condition has been explained before.
The next two conditions are of technical nature. The second one will
guarantee the boundedness of the fermionic operators. The third
imples that for $V_1, V_2\in\Se\cup\Sp\cup\Sw$ the symplectic form
$\{ V_1(v), V_2(u)\}$, and phase factors containing it linearly in
exponent, are $\Ci$ functions in both variables, bounded polynomially
in each of them. These properties make them to multipliers in the
space of Schwartz functions on $H^{\times n}$.  Finally, the fourth
condition says that a free field may be added to the cloud of the
particle.

After these preliminaries our algebra may be constructed. We
introduce formal symbols $W(V)$ for $V\in L$, $W_V$ for $V\in \Sw$,
$\dis\B_V$ for $V\in\Sp$, and $\dis\Bc_V$ for $V\in\Se$. The symbol
to which a given $V$ is attached determines the class to which it
belongs, so there is no need for special notation of $V$'s for each
case separately. Let $D$ be any finite sequence of these four symbols
and $\ch$ a Schwartz function having one four-velocity argument for
each of the symbols $\dis W_V$, $\dis\B_V$ and $\dis\Bc_V$, and one
index taking the values $\al=1,\ldots,4$, for each of the symbols
$\dis\B_V$ and $\dis\Bc_V$.  If the sequence $D$ contains $n$ symbols
$\dis W_V$, $\dis\B_V$ and $\dis\Bc_V$, and $m$ symbols $\dis\B_V$
and $\dis\Bc_V$, then $\ch\in\St(H^n, \C^{4^m})$. We introduce a new
symbol $[D](\ch)$, linear by assumption in $\ch$.  For a symbol
consisting of only one of the operators $W_V$, $\B_V$ or $\Bc_V$ the
symbolic notation is introduced
\be
W_V(\ch) = \int W_V(v)\ch(v)\, \m(v)\, ,
\label{c1}
\ee
\be
\B_V(f) = \int \B_V(v)\, \g\s v\, f(v)\, \m(v)\, ,
\label{c2}
\ee
\be
\Bc_V(f) = \int \Bc_V(v)\, \g\s v\, f(v)\, \m(v)\, ,
\label{c3}
\ee
(cf. (\ref{B}) and (\ref{Bc})) and extended by linearity to general
symbols $[D](\ch)$. The set of all formal finite sums of these
symbols forms a vector space. We divide this space by its subspace
generated by the following identifications ($G_V(v)$ is any of the
symbols $W_V(v)$, $\B_V(v)$ or $\Bc_V(v)$)
\begin{eqnarray}
&&e^{(i/2)\{ V_1, V_2\}} W(V_1) W(V_2) = W(V_1+V_2)\, ,
\label{a1}\\[3mm]
&&e^{(i/2)\{ V_1, V_2(v)\}} W(V_1) G_{V_2}(v) = G_{V_1+V_2}(v)\, ,
\label{a2}\\[3mm]
&&e^{(i/2)\{ V_2(v), V_1\}} G_{V_2}(v) W(V_1) = G_{V_1+V_2}(v)\, ,
\label{a3}\\[3mm]
&&\begin{array}{@{}l}
\dis e^{(i/2)\{ V_1(v), V_2(u)\}} W_{V_1}(v) G_{V_2}(u)\\[3mm]
\dis\hspace*{4cm}-e^{(i/2)\{ V_2(u), V_1(v)\}} G_{V_2}(u) W_{V_1}(v) =0\, ,
\end{array} 
\label{a4}\\[3mm]
&&\begin{array}{@{}l}
\dis e^{(i/2)\{ V_1(v), V_2(u)\}} 
B^{\sharp}_{V_1}(v)_\al B^{\sharp}_{V_2}(u)_\bt\\[3mm]
\dis\hspace*{4cm}+ 
e^{(i/2)\{ V_2(u), V_1(v)\}} B^{\sharp}_{V_2}(u)_\bt
B^{\sharp}_{V_1}(v)_\al =0\, ,
\end{array}
\label{a5}\\[3mm]
&&\begin{array}{@{}l}
\dis e^{(i/2)\{ V_1(v), V_2(u)\}} \B_{V_1}(v)_\al \Bc_{V_2}(u)_\bt\\[3mm]
\dis\hspace*{4cm}+
e^{(i/2)\{ V_2(u), V_1(v)\}} \Bc_{V_2}(u)_\bt \B_{V_1}(v)_\al \\[3mm]
\dis \hspace*{2cm}= \d(v, u)\,(C^{-1}\g\s v)_{\al\bt} W_{V_1+V_2}(v)\, ;
\end{array}
\label{a6}
\end{eqnarray}
if $V(v)=V_0=\con(v)$ on the support of $\ch(\ldots, v,\ldots)$ in
$v$, the variable connected with $W_V(v)$, then
\be
\lb\ldots W_V\ldots\rb (\ch) = \lb\ldots W(V_0)\ldots\rb  
\lp\int\ch(\ldots, v, \dots)\, \m(v)\rp\, .
\label{a7}
\ee
In Eq.(\ref{a5}) $\sharp={\scriptstyle 0}$ or the bar sign, the same
at both $B$'s. The phase factors appearing in (\ref{a1} -- \ref{a6})
are to be understood to multiply test functions $\ch$ in the symbols
$[D](\ch)$. The last relation says, that constant smeared Weyl
operators are identical with the standard ones. 

The elements of the factor space thus obtained will be again denoted
by $\sum_{i=1}^N [D_i](\ch_i)$ without a risk of confusion. This
vector space becomes a $*$-algebra $\sa$ with the multiplication- and
involution- law defined by 
\be
[D_1](\ch_1) [D_2](\ch_2) = [D_1D_2](\ch_1\otimes\ch_2)\, ,
\label{mult}
\ee
\be
[D](\ch)^* = [D^*](\ch^c)\, ,
\label{inv}
\ee
and the unit $\I=W(0)$.  Here $D_1D_2$ is the sequence of symbols formed
of the two sequences $D_1$ and $D_2$, $D_2$ following $D_1$. $\ch^c$
results from $\ch$ by the application of the sequence of three
operations: complex conjugation, reflection of the order of the
variables and indices, and the matrix multiplication by $C\g^0{}^
{\rm T} = - \g^0 C$ applied to each of the indices (cf. (\ref{cg})). The
sequence $D^*$ results from $D$ by reflection of its order and
subsequent replacements: $W(V)\to W(-V)$, $W_V\to W_{-V}$, $\B_V\to
\Bc_{-V}$, $\Bc_V\to\B_{-V}$. 

In the last step we consider now the problem of introducing a
$C^*$-norm on the $*$-algebra $\sa$. Let $\sn$ be the class of all
$C^*$-seminorms $p$ on $\sa$ such that \\[1mm] (i) $p([D](\ch))$ is
continuous in $\ch$ in the topology of $\St$ for each $D$;\\[1mm] (ii)
\be
p\lp W_V(\ch)\rp\leq \|\ch\|_{L^1(H, \m)}\, .
\label{wb}
\ee
A comment on each of the conditions is in place. The second one is a
necessary condition for the admitted representations of the smeared
Weyl operators $W_V(\ch)$ to be indeed given by integrals of unitary
operators with the test function $\ch$. To see the meaning of the
first condition let us compare our present context with that of the
standard algebra of the Dirac field. In the latter case one has in
the algebra the elements $[B^{\sharp_1}\ldots
B^{\sharp_n}](f^1\otimes\ldots\otimes f^n)\df B^{\sharp_1}(f^1)\ldots
B^{\sharp_n}(f^n)$, ($\sharp_i = {\scriptstyle 0}$ or the bar sign),
where $f^i\in\Sc\subset\K$.  These elements are norm continuous in
each of the functions $f^i$ in the $\St$-topology, so by the nuclear
theorem for Schwartz functions one obtains in the algebra the unique
linear extension $[B^{\sharp_1}\ldots B^{\sharp_n}](\ch)$ to the
whole of $\St(H^n, \C^{4^n})$, norm continuous in $\ch$ in the
topology of $\St$. Their analogs in our algebra are symbols
$[D](\ch)$. However, we had to define them from the beginning for the
whole space $\St$ to be able to formulate the algebraic conditions.
Condition (i) will guarantee that also here they will be continuous
extensions of products of the basic objects.

\begin{pr}
The class $\sn$ contains the maximal element $p_{max}$.  A
$C^*$-seminorm $p$ on $\sa$ is in $\sn$ iff $p\leq p_{max}$.
\label{R}
\end{pr}

The second statement is obviously true, if the first is proved.  The
proof is preceded by two lemmas.

\begin{lem}
For any $C^*$-seminorm $p\neq0$ on $\sa$ there is $p(W(V))=1$ for all
$V\in L$, and $p\lp\B_{V_1}(f)\rp = p\lp\Bc_{V_2}(f)\rp = \| f\|_\K$
for all $V_1\in\Sp$, $V_2\in\Se$ and $f\in\St$. 
\label{pB}
\end{lem}

Proof. The first statement is the consequence of the Weyl relations.
The proof of the second one is a slightly more involved version of
the analogous proof for the $C^*$-norm on the Dirac field algebra.
Let $\{ V(v), V(u)\} = F(v) - F(u)$. Then (\ref{a5}) for $V_1=V_2=V$
takes the form $$ e^{iF(v)}\B_V(v)_\al \B_V(u)_\bt +
e^{iF(u)}\B_V(u)_\bt \B_V(v)_\al =0\, .  $$ For $\ch=f\otimes g$ this
yields $\B_V(e^{iF}f)\B_V(g) + \B_V(e^{iF}g)\B_V(f) = 0$, and for
$g=f$, in particular, $\B_V(e^{iF}f)\B_V(f) =0$.  In the same way one
gets from (\ref{a6})
\be
\B_V(f)\Bc_{-V}(g) + \Bc_{-V}(e^{-iF}g)\B_V(e^{iF}f) = (f^c, g)\, .
\label{pB1}
\ee
The last two equations imply
\be
\B_V(f)\B_V(g)^*\B_V(f) = (g, f)\B_V(f)\, .
\label{pB2}
\ee
Multiplying this equation on the left by $\B_V(f)^*$ and setting
$g=f$ one gets easily $\dis p\lp \B_V(f)\rp = \| f\|_\K$ or $0$.
Assume that there is $V$ and $g\neq 0$ such that $p\lp \B_V(g)\rp
=0$. Then from (\ref{pB2}) there is $\dis |(g, f)|p\lp\B_V(f)\rp = 0$
for all $f$, hence $\dis p\lp\B_V(f)\rp = 0$ for all such $f$ that
$(g, f)\neq 0$. Each $f$ may be represented as $f=f_1+f_2$ with $(g,
f_i)\neq 0$, so $\dis p\lp\B_V(f)\rp = 0$ for all $f$, which
contradicts (\ref{pB1}) and ends the proof of the lemma. $\Box$

Let $\ch\in\St(H^n, \C^{4^m})$, $m\leq n$. There always exists a
representation $\dis\ch=\sum_{i=1}^\infty f_i^1\otimes\ldots\otimes
f_i^n$, where for a given $k$ all $f_i^k$ are either in $\Sj$ or in
$\Sc$ and the sum converges in the topology of $\St$ (e.g. the
$N$-representation \cite{ree}). There are various orders of spaces $\Sj$
and $\Sc$ in this representation possible. Denote a fixed order by
$\flat$ and the above representation with this order by
$\dis\ch=\sum_{i=1}^\infty f_i^1\otimes_\flat\ldots\otimes_\flat
f_i^n$. Let
\be
d_\flat(\ch)\df
\inf_{\ch=\sum_{i=1}^\infty f_i^1\otimes_\flat\ldots\otimes_\flat f_i^n} \,
\sum_{i=1}^\infty \|f_i^1\|_\bul \ldots\|f_i^n\|_\bul\, ,
\label{d}
\ee
where $\|f\|_\bul = \|f\|_{L^1(H, \m)}$ if $f\in\Sj$ and $\|f\|_\bul
= \|f\|_\K$ if $f\in\Sc$.

\begin{lem}
$d_\flat$ are norms on $\St(H^n, \C^{4^m})$, continuous in the
topology of $\St$.
\label{dl}
\end{lem}

Proof. We show first that $d_\flat$ is bounded by one of the
seminorms defining the topology of $\St$. We assume for simplicity
that $\ch\in\St(H^2, \C^4)$ and we are interested in the norm
$d_\flat$ for the order of spaces $(\Sj, \Sc)$. The general case
differs only by more involved notation.  Choose a Minkowski frame and
denote $\dis\ch'_\al(v, u)= v^0(u^0)^{-1/2}\sum_{\bt=1}^4
S^{-1}(u)_{\al\bt}\ch_\bt(v, u)$, where the matrix $\dis S^{-1}(u) =
(2(u^0+1))^{-1/2}(1 +\g^0\g\s u)$ induces the transformation
$\dis\fc\g\s u g= (S^{-1}(u)f)^\dagger (S^{-1}(u)g)$ (the dagger
denoting the matrix hermitian conjugation). Consider $\ch'$ as a
function of variables $\vec{v}$ and $\vec{u}$ and expand it in the
$N$-representation: $\dis\ch'_\al(v, u) = \sum_{i=1}^\infty
f_i(\vec{v}) g_{i\al}(\vec{u})$, where $f$'s and $g$'s are multiples
of products of the Hermite functions.  The sum converges in $\St$ and
$\dis \sum_{i=1}^\infty \|f_i\|_{L^1(\R^3, d^3v)}
\lp \sum_{\al=1}^4\|g_i{}_\al\|_{L^2(\R^3, d^3v)}\rp$ 
is bounded by one of the fundamental seminorms of $\ch'$, which in
turn may be bounded by one of the seminorms of $\ch$. Now it suffices
to observe that $\dis\sum_{i=1}^\infty (v^0)^{-1} f_i(\vec{v})\,
(u^0)^{1/2}
\lp S(u) g_i(\vec{u})\rp_\al$ converges to $\ch$ in $\St$ and 
$\dis\sum_{i=1}^\infty \|(v^0)^{-1} f_i(\vec{v})\|_{L^1(H,
\m)}\,\|(u^0)^{1/2}
\lp S(u) g_i(\vec{u})\rp\|_\K < \con 
\sum_{i=1}^\infty \|f_i\|_{L^1(\R^3, d^3v)}  
\lp \sum_{\al=1}^4 \|g_i{}_\al\|_{L^2(\R^3, d^3v)}\rp$ 
to get $d_\flat(\ch)<\rho(\ch)$, where $\rho$ is one of the
fundamental seminorms. The properties of a seminorm are easily
checked for $d_\flat$. That $d_\flat(\ch)=0$ implies $\ch=0$ is again
illustrated on our special case. Let $\dis\ch= \sum_{i=1}^\infty
f_i\otimes g_i$, $f_i\in\Sj$, $g_i\in\Sc$ and also choose any
$h_1\in\Sj$ and $h_2\in\Sc$. Then $$
\left|\int h_1(v)\ov{h_2(u)} \g\s u \ch(v, u)\, \m(v)\, \m(u)\right| 
\leq \sup_v |h_1(v)|\, \sum_{i=1}^\infty \|f_i\|_{L^1} \| g_i\|_\K\, .
$$ Hence, if $d_\flat(\ch) =0$ then the lhs vanishes for all $h_1$
and $h_2$ and $\ch=0$. $\Box$

Proof of Proposition \ref{R}.  Let $\flat(D)$ be the order of
spaces $\Sj$ and $\Sc$ corresponding to the order of symbols $W_V$
and $B_V^\sharp$, respectively, in the sequence $D$.  Then by the
assumption (\ref{wb}) and Lemma \ref{pB} one has for any $p\in\sn$ 
$$
p\lp [D](f^1\otimes_{\flat(D)}\ldots\otimes_{\flat(D)}f^n)\rp \leq
\|f^1\|_\bul \ldots\|f^n\|_\bul\, .
$$ 
If $\dis\ch= \sum_{i=1}^\infty
f_i^1\otimes_{\flat(D)}\ldots\otimes_{\flat(D)}f_i^n$ then by the
assumed continuity $\dis p([D](\ch))\leq \sum_{i=1}^\infty 
\|f_i^1\|_\bul \ldots\|f_i^n\|_\bul$. Hence for any element of $\sa$ one has 
$$ 
p\lp\sum_{k=1}^N [D_k](\ch_k)\rp \leq \sum_{k=1}^N
d_{\flat(D_k)}(\ch)\, .  
$$ 
We define the seminorm $p_{{\rm max }}$ on
$\sa$ by $$ p_{{\rm max}}(A)= \sup_{p\in\sn} p(A)\, .  $$ There is
$p_{{\rm max}}\lp[D](\ch)\rp \leq d_{\flat(D)}(\ch)$, so $p_{{\rm max
}}\in\sn$. $\Box$

The answer to the question whether $p_{{\rm max }}$ is a norm on $\sa$
is not known yet. If it is not, one divides $\sa$ through the ideal
$\Id$ of those elements for which $p_{{\rm max }}$ vanishes. The
seminorm $p_{{\rm max }}$ induces then a $C^*$-norm $\|.\|$ on $\sa
/\Id$ in the standard way. The completion of $\sa /\Id$ in this norm
is a $C^*$-algebra $(\Fa, \|.\|)$. We propose to regard this algebra
as the base of a theory of asymptotic fields.  With regard to the
interpretation of electromagnetic ingredients of the algebra, one
should have in mind the remarks made at the end of previous section. 

It is easy to see that representations of the algebra $(\Fa, \|.\|)$
are in natural $1:1$ correspondence with those representations $\pi$
of $\sa$ for which $p(.)=\|\pi(.)\|\in\sn$. 

In the asymptotic algebra of fields there is no place more for the
local gauge transformations. The only gauge dependent quantity in the
electromagnetic test fields is the additive constant in $\F(l)$. This
freedom is closely connected with the global gauge transformation of
the charge carrying fields, which is implemented in the algebra
itself. Let $W(V)=W(0, c)$, $c=\con$ (i.e. $V={\rm pure~gauge}$,
$\F=c$), and set $\g_c(A) = W(0, c) A\, W(0, c)^*$. Then $\g_c(A) =A$
for $A=W(V)$ or $W_V(\ch)$, and $\dis\g_c(\B_V(f))= e^{-ice}\B_V(f)$,

$\dis\g_c(\Bc_V(f))= e^{+ice}\Bc_V(f)$. Algebra $\sa$ is the linear
span of its subspaces $\sa_k$, $k\in\Z$, where $\dis\g_c= e^{icke}\,
{\rm id}$ on $\sa_k$. 
The subspace $\sa_0$ is a $*$-subalgebra of $\sa$. If $\Id\neq 0$ then
it is easily seen that $\Id$ is the linear span of $\Id_k\df\sa_k\cap
\Id$. The 
decomposition is therefore inherited by $\Fa$, and $\Fa_0$ is a
$C^*$-algebra, which may be interpreted as the algebra of observables
$\Ob\equiv\Fa_0$.

The algebraic relations of $\sa$ have been obtained by treating
$B_V^\sharp(v)$ heuristically as products of $B^\sharp(v)$ and Weyl
operators for charged fields $W(V(v))$. In the next section we shall
see that this heuristic idea may be also used to obtain a class of
representations of $\sa$ (and $\Fa$).

\setcounter{equation}{0}
\setcounter{pr}{0}

\section{A class of representations}
\label{rep}

Let $W_0(V)$ be a representation in a Hilbert space $\Hi_1$ of the
Weyl algebra over the test function space $\hat{L}$ with the
symplectic form (\ref{sn}). We assume that for any $V(.)\in\Se$ it
satisfies the following conditions\\ (i) for every $\ph\in\Hi_1$ the
vectors $W_0 (V(v))\ph$, $v\in H$, span a separable subspace;\\ (ii)
for every $\ph$, $\psi\in\Hi_1$ the function $(\ph, W_0(V(v))\psi)$ is
measurable in $v$.\\ (The class of representations satifying the
conditions is nonempty -- this may be shown by an explicit
construction making use of the usual Fock representation and one of
representations discussed in \cite{her94}.) It follows that $(\ph,
W_0(V_1(v_1))\ldots W_0(V_n(v_n))\psi)$ is also measurable. For
$V_i\in\Se\cup\Sp\cup\Sw$, $i=1,\ldots ,n$, $\ch\in L^1(H^n, \m^n)$
we denote
\begin{eqnarray*}
&&\lb W_0{}_{V_1}\ldots W_0{}_{V_n}\rb(\ch)= \\[1mm] &&\int
W_0(V_1(v_1))\ldots W_0(V_n(v_n))\ch(v_1,\ldots, v_n)\,
\m(v_1)\ldots\m(v_n)\, ,
\end{eqnarray*}
the integral in the weak sense. The Weyl algebra relations imply
\be
\lb W_0{}_{V_1} W_0{}_{V_2}\rb(\chi) 
=\lb W_0{}_{V_2} W_0{}_{V_1}\rb(\chi')\, ,
\label{wcom}
\ee
where $\ch'(u, v) = e^{-i\{V_1(v), V_2(u)\} }\ch(v, u)$. If, in
particular, $\{V_1(v), V_2(u)\} = F_{12}(v) - F_{12}(u)$, then 
\be
W_0{}_{V_1}(\ch_1) W_0{}_{V_2}(\ch_2) = W_0{}_{V_2}(e^{iF_{12}}\ch_2)
W_0{}_{V_1}(e^{-iF_{12}}\ch_1)\, .
\label{wvf}
\ee

Let further $B(f)$ (and $B^\sharp(f)$) be a concrete realization
(representation) of the Dirac field algebra in a Hilbert space
$\Hi_2$.  We would like to define $\B_V(v)$ as a product of
$W_0(V(v))$ and $\B(v)$, that is to give sense to the expression 
$$
\B_V(f) = \int W_0(V(v))\otimes\B(v)\, \g\s v \, f(v)\, \m(v) 
$$ 
for $f\in\Sc$. (When constructing the representation we shall use
the simplified notation $\B_V(f)$ instead of the more appropriate
$\pi(\B_V(f))$ etc., which may be restored at the end of
construction.) Let $\{ e_i\}$ be an orthonormal basis of the Hilbert
space $\K$.  If we "expand" $\B(v)$ in this basis, we are led to the
following formulation. Consider a family of bounded operators 
on $\Hi_1\otimes\Hi_2$ 
$$
\B_V(f)_n \df \sum_{i=1}^n W_0{}_V(\ov{e_i}\G f)\otimes \B(e_i)\, ,
$$ 
where $(\G f)(v) = \g\s v f(v)$. 

\begin{pr}
The sequence of operators $\B_V(f)_n$ converges $*$-strongly to a
bounded operator $\B_V(f)$, with $\|\B_V(f)\|\leq\|f\|_\K$. 
\label{limb}
\end{pr}

Proof. Let $\{ V(v), V(u)\} = F(v) - F(u)$. From the Dirac field
anticommutation relations and (\ref{wvf}) we obtain
\begin{eqnarray*}
&&\lp\B_V(f)_n - \B_V(f)_m\rp^* \lp\B_V(f)_n - \B_V(f)_m\rp \\[2mm]
&&+ \lp\B_V(e^{-iF}f)_n - \B_V(e^{-iF}f)_m\rp
\lp\B_V(e^{-iF}f)_n - \B_V(e^{-iF}f)_m\rp^* \\[1mm]
&&\hspace*{2cm}= \lp\sum_{i=m+1}^n w_i^* w_i\rp\otimes\I\, ,
\end{eqnarray*}
where $w_i= W_0{}_V(\ov{e_i}\G f)$. For any vector
$\psi\in\Hi_1\otimes\Hi_2$ there is then
\be
\begin{array}{@{}l}
\dis\|\lp\B_V(f)_n - \B_V(f)_m\rp\psi\|^2 + 
\|\lp\B_V(e^{-iF}f)_n^* - \B_V(e^{-iF}f)_m^*\rp\psi\|^2 \\[2mm]
\hspace*{2cm}\dis =p_\psi\lp \sum_{i=m+1}^n w_i^* w_i\rp\, ,
\end{array}
\label{cau}
\ee
where $p_\psi(A)\df(\psi, A\otimes\I\,\psi)$ agrees for positive $A$
with one of the seminorms defining the $\si$-weak topology on the
space of bounded operators on $\Hi_1$. We shall show below that 
\be
\sum_{i=1}^\infty w_i^* w_i= \| f\|^2\,\I\, ,
\label{sig}
\ee
and that the series converges $\si$-strongly. Hence $\B_V(f)_n$ and
$\B_V(e^{-iF}f)_n^*$ converge strongly to bounded operators for all
$f$, which implies the $*$-strong convergence of $\B_V(f)_n$ for all
$f$.  Putting $m=0$ in (\ref{cau}) and taking the limit in $n$ we
obtain the bound of the norm. To prove (\ref{sig}) observe first that
for any $x$, $y\in \Hi_1$ we have $\dis (y, w_i\, x)= \int (y,
W_0(V(v))\, x)\,\ov{e_i(v)}\g\s v f(v)\, \m(v) = (e_i, (y,
W_0(V(.))x)\, f)_\K$, so that $$
\sum_{i=1}^\infty |(y, w_i\, x)|^2 = \|(y, W_0(V(.))x)f\|^2_\K\, .
$$ For fixed $x$ let $\{ \ph_j\}$ be an orthonormal basis of the
subspace of $\Hi_1$ spanned by $W_0(V(v))x$, $v\in H$. Then 
\begin{eqnarray*}
&&\sum_{i=1}^\infty (x, w_i^* w_i\, x) = \sum_{i,j=1}^\infty |(\ph_j,
w_i\, x)|^2\\[2mm] &&=\sum_{j=1}^\infty \int |(\ph_j,
W_0(V(v))x)|^2\,\ov{f(v)}\g\s v f(v)\, \m(v) = \|f\|^2_\K\,\|x\|^2\,
,
\end{eqnarray*}
the last equality by the Lebesgue theorem.  As $\sum_{i=1}^n w_i^*
w_i$ is an increasing sequence of positive operators, the above
calculation shows that (\ref{sig}) holds in the $\si$-strong sense
(e.g. \cite{bra}, Lemma 2.4.19). $\Box$

The building blocks of the representation acting on 
$\Hi_1\otimes\Hi_2$ are now defined by
\begin{eqnarray}
&&W(V)= W_0(V)\otimes\I\, ,~~~~V\in L\, ,
\label{rep1}\\[2mm]
&&W_V(\ch)= W_0{}_V(\ch)\otimes\I\, ,~~~V\in\Sw\, ,~~ \ch\in\Sj\, ,
\label{rep2}\\[2mm]
&&\begin{array}{@{}l}
\dis B_V^\sharp(f) =\sum_{i=1}^\infty W_0{}_V(\ov{e_i}\G f)
\otimes B^\sharp (e_i)\, ,\\[2mm]
\hspace*{4cm}\dis V\in\Se\cup\Sp\, , f\in\Sc\, .
\end{array}
\label{rep3}
\end{eqnarray}
The definition of $B_V^\sharp$ is independent of the choice of the
basis $\{ e_i\}$, as it is easily shown that 
\be
(x_1\otimes y_1, B_V^\sharp(f)\, x_2\otimes y_2) =
\lp y_1, B^\sharp((x_1, W_0(V(.))x_2)f)\, y_2\rp\, .
\label{rep4}
\ee
If $D$ is any sequence of the symbols $W(V)$, $W_V$ and $B_V^\sharp$,
and $(f^1,\ldots , f^n)$ form a sequence of type $\flat(D)$, then we
define $[D](f^1\otimes\ldots\otimes f^n)$ as the product of the
"building blocks".  In view of Prop.\ref{limb} and of the obvious
bound $\|W_V(\ch)\|\leq\|\ch\|_{L^1}$, this element is norm
continuous in each of $f$'s in the topology of $\St$. By the nuclear
theorem it extends then to the function $[D](\ch)$ norm continuous in
$\ch$ in the $\St$-topology.  The conditions (\ref{mult}) and
(\ref{inv}) are satisfied, and the operator norm fulfills the
defining conditions of the class $\sn$. To complete the proof that we
have thus obtained a representation of the algebra $\Fa$ it remains
to show that the relations (\ref{a1}--\ref{a7}) are satisfied.

The conditions (\ref{a1}) and (\ref{a7}) are obviously satisfied. It
is sufficient to check the other relations for elements $[D](\ch)$
with $D$'s being sequences of two symbols. The relations (\ref{a2})
and (\ref{a3}) are then quite obvious as well. Eq. (\ref{a4}) for
$G_V=W_V$ follows from (\ref{wcom}). For $G_V=B_V^\sharp$ it is easy
to show that
\begin{eqnarray*}
&&(x_1\otimes y_1, \lb W_{V_1} B_{V_2}^\sharp \rb(\ch)\, x_2\otimes
y_2)\\[2mm] &&= \lp y_1, B^\sharp\lp\int (x_1, W_0(V_1(v))W_0(V_2(.))
x_2)\,\ch(v, .)\,
\m(v) \rp y_2\rp\, ,
\end{eqnarray*}
and similarly in the opposite order of symbols, which implies
(\ref{a4}).

To prove (\ref{a5}) and (\ref{a6}) we have to make a digression on
the extension of products $B^{\sharp_1}(f) B^{\sharp_2}(g)$ in the
algebra of the Dirac field. We have mentioned such an extension to
the space of Schwartz functions, but now we need a wider family. 

Let $\K\otimes\K$ be the tensor product Hilbert space. This space
consists of measurable functions $\ch_{\al\bt}(v, u)$, for which 
$$
\sum_{\vspace*{-2mm}\begin{array}{l}
\scriptstyle \al, \al'\\[-2mm]
\scriptstyle \bt, \bt'
\end{array}} \int \ch_{\al\bt}^*(v, u)(\g^0\g\s v)_{\al\al'} 
(\g^0\g\s u)_{\bt\bt'}
\ch_{\al'\bt'}(v, u)\, \m(v)\,\m(u) < \infty\, .
$$ 
Let, further, $\K\otimes_1\K$ be the subspace of $\K\otimes\K$
consisting of those $\ch\in\K\otimes\K$ for which $$
\|\ch\|_1\df\inf_{\ch=\sum_{i=1}^\infty f_i\otimes g_i} 
\, \sum_{k=1}^\infty \|f_k\| \|g_k\| <\infty\, .
$$ $\|.\|_1$ is a norm on $\K\otimes_1\K$, $\|\ch\|<\|\ch\|_1$ for
$\ch\in\K\otimes_1\K$, and $(\K\otimes_1\K, \|.\|_1)$ is a Banach
space.  These statements follow most simply from the following two
observations. \\ (i) $(\K\otimes\K, \|.\|)$ is isomorphic with the
space of Hilbert-Schmidt operators on $\K$ by the map
$\ch\to\Op_\ch$, where for $\ch=\sum_{i=1}^\infty f_i\otimes g_i$ the
operator $\Op_\ch$ is defined by
\be
\Op_\ch h = \sum_{i=1}^\infty (h^c, f_i) g_i\, .
\label{och}
\ee
(ii) Under the same map $(\K\otimes_1\K, \|.\|_1)$ is isomorphic with
the Banach space of trace class operators on $\K$ \cite{sch}. Thus, if
$\ch_n\to\ch$ in $\K\otimes_1\K$ then $\ch_n\to\ch$ in $\K\otimes\K$.

Extension of the product of fundamental elements in the Dirac field
algebra is now easily achieved. For $\dis\ch=\sum_{i=1}^n f_i\otimes
g_i$, $f_i,\, g_i\in\K$, we set $\dis [B^{\sharp_1}B^{\sharp_2}](\ch)
= \sum_{i=1}^n B^{\sharp_1}(f_i)B^{\sharp_2}(g_i)$. This defines a
linear map of the algebraic product $\Ka$ (densly contained in
$\K\otimes_1\K$) into the algebra, with the norm bound
$\dis\|[B^{\sharp_1}B^{\sharp_2}](\ch)\|\leq\|\ch\|_1$. Hence the map
extends to the whole $\K\otimes_1\K$, with the conservation of the
bound.  For $\ch\in\St$ this reduces to the extension mentioned
previously.

The anticommutation relations may be now extended from $\Ka$ to the
whole $\K\otimes_1\K$, which gives 
\begin{eqnarray}
&&[B^\sharp B^\sharp](\ch +\ch^{{\rm T}}) = 0\, ,
\label{ecom1}\\[2mm]
&&[\B\Bc](\ch) + [\Bc\B](\ch^{{\rm T}}) = {\rm Tr} \Op_\ch \,\I\, ,
\label{ecom2}
\end{eqnarray}
where $\ch^{{\rm T}}{}_{\al\bt} (v, u)= \ch_{\bt\al}(u, v)$, and
${\rm Tr}\Op_\ch$ is the trace of the operator (\ref{och}),
$\dis{\rm Tr}\Op_\ch = \sum_{i=1}^\infty (f^c_i, g_i)$ for
$\dis\ch=\sum_{i=1}^\infty f_i\otimes g_i$.

After these preparations we take up the proof of the relations
(\ref{a5}) and (\ref{a6}). For $x_1,\, x_2\in\Hi_1$ let $\ph_j$ be an
orthonormal basis of the linear span of vectors $W_0(V_1(v))^* x_1$
and $W_0(V_1(v))x_2$, $v\in H$. Let, further, $\psi_\la$ be an
orthonormal basis (not necesserily countable) of $\Hi_2$, and $f,\,
g\in\Sc$.  Expanding $B_{V_2}^{\sharp_2} (g) x_2\otimes y_2$ in the
basis $\ph_j\otimes\psi_\la$ and using (\ref{rep4}) one finds 
\begin{eqnarray*}
&&\lp x_1\otimes y_1, B_{V_1}^{\sharp_1} (f) B_{V_2}^{\sharp_2} (g)
x_2\otimes y_2\rp \\[2mm] &&= \sum_{i=1}^\infty \lp y_1,
[B^{\sharp_1}B^{\sharp_2}]
\lp(x_1, W_0(V_1(.))\ph_i)f\otimes (\ph_i, W_0(V_2(.))x_2)g\rp y_2\rp\, .
\end{eqnarray*}
We have
\begin{eqnarray*}
&&\sum_{i=1}^\infty
\|(x_1, W_0(V_1(.))\ph_i)f\|_\K\, \|(\ph_i, W_0(V_2(.))x_2)g\|_\K \\
&&\leq\lp\sum_{i=1}^\infty\|(W_0(V_1(.))^* x_1,
\ph_i)f\|_\K^2\rp^{1/2}
\lp\sum_{i=1}^\infty\|(\ph_i, W_0(V_2(.))x_2)g\|_\K^2\rp^{1/2}\\[2mm]
&&= \| x_1\| \| x_2\| \|f\|_\K\|g\|_\K\, ,
\end{eqnarray*}
the last equality by the Lebesgue theorem. Hence, the series
\be
\sum_{i=1}^\infty (x_1, W_0(V_1(.))\ph_i)f\otimes (\ph_i, W_0(V_2(.))x_2)g
\label{psi}
\ee
converges both in $\K\otimes_1\K$ and $\K\otimes\K$, to the same
element.  The limit in $\K\otimes\K$ is easily found as the point
limit of functions, which yields $(x_1, W_0(V_1(v)) W_0(V_2(u)) x_2)
f(v)g(u)$. Thus we have proved that if $\ch\in\Ka$ then 
\be
\begin{array}{@{}l}
\dis (x_1, W_0(V_1(.)) W_0(V_2(.)) x_2)\, \ch\in\K\otimes_1\K\, ,\\[3mm]
\dis \|(x_1, W_0(V_1(.)) W_0(V_2(.)) x_2)\, \ch\|_1\leq\| x_1\| \| x_2\| 
\|\ch\|_1\, ,
\end{array}
\label{at}
\ee
and for $\ch\in\Sc\otimes_{{\rm alg}}\Sc\subset\Ka$
\be
\begin{array}{@{}l}
\dis\lp x_1\otimes y_1, [B_{V_1}^{\sharp_1} B_{V_2}^{\sharp_2}](\ch) 
\, x_2\otimes y_2\rp \\[3mm]
\dis =\lp y_1, [B^{\sharp_1}B^{\sharp_2}] 
\lp(x_1, W_0(V_1(.)) W_0(V_2(.))x_2)\ch\rp\, y_2\rp\, .
\end{array}
\label{red}
\ee
By continuity (\ref{at}) remains true for $\ch\in\K\otimes_1\K$, and
(\ref{red}) for $\ch\in\St(H^2, \C^{4^2})\subset\K\otimes_1\K$.
Denote $\dis\ch'_{\al\bt}(v, u)=e^{(i/2)\{ V_1(v), V_2(u)\}
}\ch_{\al\bt}(v, u)$ and $\dis\ch''_{\al\bt}(v, u)=e^{(i/2)\{ V_2(u),
V_1(v)\} }
\ch^{{\rm T}}_{\al\bt}(v, u)$. Using the Weyl relations for $W_0$ one obtains 
from (\ref{red})
\begin{eqnarray*}
&&\lp x_1\otimes y_1, \lp [B_{V_1}^{\sharp_1}
B_{V_2}^{\sharp_2}](\ch') + [B_{V_2}^{\sharp_2}
B_{V_1}^{\sharp_1}](\ch'')\rp x_2\otimes y_2\rp \\[3mm] &&=\lp y_1,
\lp [B^{\sharp_1} B^{\sharp_2}](\Psi) + 
[B^{\sharp_2} B^{\sharp_1}](\Psi^{{\rm T}})\rp y_2\rp\, ,
\end{eqnarray*}
where $\dis\Psi_{\al\bt}(v, u) = \lp x_1,
W_0(V_1(v)+V_2(u))x_2\rp\ch_{\al\bt}(v, u)$. The relation (\ref{a5})
follows now from (\ref{ecom1}). The proof of (\ref{a6}) will be
complete by (\ref{ecom2}) and the definition (\ref{rep2}) if we show
that $$
{\rm Tr} \Op_\Psi = \int\lp x_1, W_0(V_1(v)+V_2(v))\, x_2\rp \,\sum_{\al,\bt}
(C^{-1}\g\s v)_{\al\bt}\, \ch_{\al\bt}(v, v)\, \m(v)\, .  $$ The rhs may
be written as $$
\int\lp x_1, W_0(V_1(v))W_0(V_2(v))\, x_2\rp \,\sum_{\al,\bt}
(C^{-1}\g\s v)_{\al\bt}\,\ch'_{\al\bt}(v, v)\, \m(v)\, .  
$$ 
This formula
defines a distribution on $\ch'\in\St$, so it is sufficient to take
$\ch'=f\otimes g$, $f,\,g\in\St(H, \C^4)$. Then $\Psi$ is given by
(\ref{psi}) and 
\begin{eqnarray*}
&&{\rm Tr}\Op_\Psi = \sum_{i=1}^\infty \lp \ov{(x_1,
W_0(V_1(.))\ph_i)} f^c, (\ph_i, W_0(V_2(.))x_2) g\rp\\
&&=\sum_{i=1}^\infty \int (x_1, W_0(V_1(.))\ph_i)\, (\ph_i,
W_0(V_2(.))x_2)
\,\ov{f^c(v)}\g\s v g(v)\, \m(v)\, ,
\end{eqnarray*}
which yields the desired relation by the Lebesgue theorem. This ends
the proof of the conditions of our algebra.

\setcounter{equation}{0}
\setcounter{pr}{0}

\section{Discussion and outlook}
\label{dis}

We have shown how heuristic quantization of the asymptotic structure
of classical field electrodynamics leads to the construction of an
asymptotic algebra of fields, whose states may be expected to
describe the structure of collision states in quantum
electrodynamics, including the charged states. This algebra is a
$C^*$-algebra, so there is no need for the indefinite metric
formalism. The charged fields are accompanied by Coulomb fields,
which solves the problem of Gauss' law in charged states. The
construction depends on the choice of a class of "satellite" fields.
This choice was left open to some extent.  It corresponds to
selecting various "clouds" of free radiation field accompanying the
particle in addition to the Coulomb field. Three classes of satellite
fields satisfying the defining conditions (i)--(iv) in Sec.\ref{alg}
are worth mentioning:\\ (a) the class of all fields satisfying the
conditions (i)--(iv);\\ (b) the subclass consisting of fields of the
form $V_a(v; s, l) =V_e{}_a(v, l) + V_0{}_a(s, l)$, where $V_e{}_a(v,
l)$, defined in (i), corresponds to the Coulomb field, and
$V_0{}_a\in L$;\\ (c) the subclass of fields for which $V_a(v;
-\infty, l)$ does not depend on $v$.\\ The first choice is the most
general one within the limits of our construction.  The second
possibility is the simplest one. In that case the smeared Weyl
elements reduce to the (simple) Weyl elements, as the space $\Sw$ is
then naturally isomorphic with $L$. We mention as an aside that an
explicit faithful representation of the corresponding algebra can be
constructed. The choice (c) has a clear physical interpretation: the
long-range tail of the cloud accompanying the particle is chosen to
compensate the velocity dependence of the tail of the Coulomb field.
The total satellite field has then a velocity independent flux at
spatial infinity.

The last possibility seems to be the one closest to the picture
emerging from the analysis of superselection sectors structure in the
algebraic framework of local observables. However, we leave the
problem of specifying the physically justified choice of satellite
fields open at present.

The solution of this problem may depend on the answer to the
important question of whether our algebra may be obtained by some
limiting procedure from the field algebra of the full theory. This
seems a difficult problem, but the classical theory gives hints, how
such a limiting process could look like. The quantum version of the
null infinity limit of the electromagnetic field may be thought of as
an LSZ-type limit in lightlike directions, which brings in mind the
construction of Buchholz \cite{buch82}. 
For the matter field one would have to choose a gauge in a class 
supplying a quantum analog of the class mentioned in Sec.\ref{clas}.
An LSZ-type limit on the hyperboloid $x^2=\la^2$, $x^0>0$ with
$\la\to\infty$ may then be expected to exist.

Important as the latter problem may be, the asymptotic algebra also
deserves further investigations on its own. The following physically
interesting problems may be posed within this framework. \\ (i) How
can the local observables be characterized and what is their relation
to the nonlocal ones? That the latter observables are present may be
read off from the properties of the symplectic form.\\ (ii) The
Poincar\'e group acts naturally on the algebra as a group of
automorphisms. Do there exist irreducible Poincar\'e-covariant
representations of the algebra? If so, which superselection structure
of the algebra of local observables is implied? (We recall that in
charged superselection sectors the Lorentz group has to be
spontaneously broken). \\ (iii) What is the measure class of the
spectrum of energy-momentum in translation-covariant, positive energy
representations (infraparticle problem)?

Finally, let us mention for completeness that the construction given
in this paper may be reflected in time, yielding the asymptotic
"in"-algebra.  If these two algebras fit into the interacting theory,
the scattering problem may be considered. One may hope, for instance,
that the perturbation calculus in a suitable gauge, starting from the
quasi-free theory supplied by our algebra, should be
infrared-regular.

\section*{Acknowledgements}
\label{ack}
I would like to thank Professor D.\,Buchholz for his steady interest
in my work. I have profited greatly from the many discussions which we
had.  A discussion with Professor R.\,Haag is also gratefully
acknowledged.  I am grateful to the II. Institut f\"{u}r Theoretische
Physik, Universit\"{a}t Hamburg, for hospitality and to the Humboldt
Foundation for financial support.

\setcounter{equation}{0}
\setcounter{pr}{0}
\setcounter{section}{0}
\renewcommand{\thesection}{Appendix}
\renewcommand{\theequation}{\Alph{section}.\arabic{equation}}
\renewcommand{\thepr}{\Alph{section}.\arabic{pr}}

\section{Equivalence of spinor and tensor formulas}
\label{app}

We prove here the equivalence of tensor and spinor versions of those
formulas which appear in Sec.\ref{clas}, but were given only in the
spinor form in \cite{her95}.

For $l_a=o_Ao_{A'}$ in the notation of \cite{her95}, $\D_A=\D/\D o^A$
and $\D_a=\D/\D l^a$, one has $\D_A\al(l) = o^{A'}\D_{AA'}\al(l)$,
hence
\be
(l_a\D_b - l_b\D_a)\al(l) = -(\e_{A'B'}o_{(A}\D_{B)} +
\e_{AB}o_{(A'}\D_{B'})
\al(l)\, ,
\label{adif}
\ee
and the integral identity (\ref{ip}) is then equivalent to (A8) of
\cite{her95}.

The electromagnetic field tensor and spinor are connected by $F_{ab}
= \e_{A'B'}\ph_{AB} + \e_{AB}\ov{\ph}_{A'B'}$, hence the field
$\varrho_{AA'}(x)=\varphi_{AB}(x) x^B_{A'}$ ((2.28) in \cite{her95}) is
equivalently expressed as $\varrho_a(x)=x^b \, {}^-F_{ba}(x)$, where
${}^-F_{ba}$ is the anti-selfdual part of $F_{ba}$. Its null
asymptotic $\lim_{R\to\infty} R \varrho_a(x+Rl)= N_a(x\s l, l)$ is
given by $N_a(s, l)=\D_{A'}\z_A(s, l)$, where $\z_A(s,
l)=o_{A'}V^{A'}_A$, $V_a(s, l)$ defined in (\ref{null}) (see
\cite{her95}, Eq.(2.44)). We have 
$$
\D_{A'}\z_A= -V_a+o_{B'}\D_{A'}V^{B'}_A = -V_a-\frac{1}{2}\e_{B'A'} 
o^{C'}\D_{C'}V^{B'}_A +o_{(B'}\D_{A')}V^{B'}_A\, .  $$ In view of
homogeneity (\ref{hom}) and using (\ref{adif}) this may be written as

$$ N_a(s, l)=\frac{1}{2}(s\dV_a(s, l)-V_a(s,
l))+{}^+(l_a\D_b-l_b\D_a)V^b(s, l)\, , $$ where
${}^+(l_a\D_b-l_b\D_a)$ is the selfdual part of $l_a\D_b-l_b\D_a$. It
was shown in \cite{her95} (and may be also shown without use of spinors)
that from (\ref{hom}) and (\ref{Q}) now follows that the limit values

$N_a(\pm\infty, l)$ are proportional to $l_a$, that is
\begin{eqnarray*}
&&-\frac{1}{2}V_a(+\infty, l) +{}^+(l_a\D_b-l_b\D_a)V^b(+\infty, l)
=-l_a q(l)\, ,\\ &&-\frac{1}{2}V_a(-\infty, l)
+{}^+(l_a\D_b-l_b\D_a)V^b(-\infty, l) =-l_a \k(l)\, ,
\end{eqnarray*}
which corresponds with the equations (2.54) and (2.56) of \cite{her95}
(there we used $\si=\k-q$ instead of $\k$). The conditions of reality
of $q$ and $\k$ ((3.32) in \cite{her95}), and the above equations in
that case are now equivalent to (\ref{cond}), and (\ref{q}) and
(\ref{k}) respectively.

It was shown in \cite{her95} (Eq.(2.64)) that
\be
o^{A'}\Vo{}_{A'A}(-\infty, l) = \D_A\F(l)
\label{gs}
\ee
for some $\F(l)$ homogeneous of degree $0$. In view of (\ref{adif})
this is equivalent to (\ref{gra}). The gauge of $\Vo{}_a(-\infty, l)$
(\ref{g}) is written with the use of (\ref{adif}) as $$
\Vo{}_a(-\infty, l)=\D_A(o^B g_{BA'}(l))+\D_{A'}(o^{B'}g_{B'A}(l))\, .
$$ As $o^B g_{BA'}(l)$ are in $1:1$ correspondence with $G_{ab}$, the
gauge has the form $\Vo{}_a(-\infty, l)=\D_A h_{A'}(o, \oc)+\D_{A'}
\ov{h}_A(o, \oc)$, 
where $h_{A'}(\al o, \ac\oc)=\ac^{-1} h_{A'}(o, \oc)$, $h_{A'}(o,
\oc)o^{A'}=\F(l)$, and in the statements (i)--(iii) following 
(\ref{g}) the tensor $G_{ab}$ may be replaced by $h_{A'}(o, \oc)$
satisfying these conditions. This representation satisfies
(\ref{gs}), so (i) is proved. We get a special gauge choosing $$
h_{A'}(o, \oc) = \frac{t_{A'A}o^A}{t\s l}\F(l)\, , $$ where $t$ is
any unit, positive timelike vector. Any other gauge differs by
$\bt(l)l_a$, $\bt(l)$ homogeneous of degree $-2$. Let
$\al(l)=\bt(l)-c\,(t\s l)^{-2}$, with such a constant $c$ that
$\int\al(l)\, \dl=0$. There exists then a homogeneous of degree~$0$
function $A(l)$, such that $\D_A\D_{A'}A(l)=\frac{1}{2} o_Ao_{A'}
\al(l)$. 
$A(l)$ is determined up to an additive constant. The new gauge is
then determined by $$ h'_{A'}(o, \oc) = \frac{t_{A'A}o^A}{t\s
l}(\F(l)+c) + \D_{A'}A(l)\, .  $$ With this formula the statements
(ii) and (iii) following (\ref{g}) are seen to be true. 

Finally, the angular momentum term $\Delta M_{ab}$ (\ref{am}) is the
tensor version of the angular momentum spinor term $\dis\Delta
\mu_{AB}
=\frac{1}{2\pi}\int qo_{(A}\D_{B)}\F\, \dl$, easily obtained from
$\Delta M_{ab}= \e_{A'B'}\Delta\mu_{AB} + \e_{AB}
\ov{\Delta\mu}_{A'B'}$ by
the use of (\ref{adif}).

\end{sloppypar}
\end{document}